\documentclass[12pt,aps,prd,showpacs,nofootinbib,preprintnumbers,a4paper]{revtex4-1}

\usepackage{amsmath}
\usepackage{comment}
\usepackage[dvipdfmx]{graphicx}

\usepackage{dcolumn}
\usepackage{bm}
\usepackage[dvipdfmx]{color}
\usepackage[normalem]{ulem}

\allowdisplaybreaks

\newcommand{\nn}{\nonumber}

\newcommand{\be}{\begin{eqnarray}}
\newcommand{\ee}{\end{eqnarray}}
\catcode`\@=11
\def\lsim{\mathrel{\mathpalette\@versim<}}
\def\gsim{\mathrel{\mathpalette\@versim>}}
\def\@versim#1#2{\vcenter{\offinterlineskip
\ialign{$\m@th#1\hfil##\hfil$\crcr#2\crcr\sim\crcr } }}
\catcode`\@=12
\newcommand{\del}{\partial}

\newcommand{\Slash}[1]{{\ooalign{\hfil#1\hfil\crcr\raise.167ex\hbo x{/}}}}

\def\thefootnote{\fnsymbol{footnote}}

\begin{document}

\title{
Scale and electroweak first-order phase transitions}

\author{Jisuke \surname{Kubo}}
\email{jik@hep.s.kanazawa-u.ac.jp}
\affiliation{Institute for Theoretical Physics, Kanazawa University, Kanazawa 920-1192, Japan}

\author{Masatoshi \surname{Yamada}}
\email{masay@hep.s.kanazawa-u.ac.jp}
\affiliation{Institute for Theoretical Physics, Kanazawa University, Kanazawa 920-1192, Japan}

\preprint{KANAZAWA-15-09}

\pacs{11.30.Ly,11.15.Tk,12.60.Rc,95.35.+d }

\begin{abstract}
We consider phase transitions in
the standard model  (SM) without the Higgs mass term,
 which is coupled through a Higgs portal term
to an SM singlet,
 classically scale-invariant gauge sector
with  SM singlet scalar fields.
At lower energies
the gauge-invariant scalar bilinear in the hidden sector forms
a condensate, dynamically creating a robust energy scale, which
is transmitted through the  Higgs portal term to the SM sector.
A scale phase transition  is  a transition between phases
with  zero and nonzero condensates.
An interplay between the EW and scale phase transitions is therefore expected.
We find that in a certain parameter space  both 
the electroweak (EW) and scale phase transitions can be a strong first-order phase transition.
The result is obtained 
by means of an effective theory for the condensation
of scalar bilinear in the mean field approximation.

 \end{abstract}
\setcounter{footnote}{0}
\def\thefootnote{\arabic{footnote}}
\maketitle

\section{Introduction}
\noindent Thanks to the recent discovery of the Higgs boson at LHC~\cite{Aad:2012tfa,Chatrchyan:2012ufa}
the standard model (SM) describing the dynamics of elementary particles is 
now complete.
However, the SM  accommodates neither dark  matter (DM) 
nor neutrinos with a finite mass.
Therefore, the SM is incomplete as a theory to explain phenomena in our Universe,  and consequently it
 has to be  extended.
These unsatisfactory features are the main motivations for probing 
both theoretically and experimentally new physics around the TeV  scale. 

Besides the problems mentioned above there are also 
problems of a more theoretical nature.
One of them is the origin of the electroweak (EW) scale.
Certainly, the SM cannot explain it, but a hint might exist in the SM:
The Higgs mass term is the only term that breaks scale invariance at the classical level. 
In fact there have recently been many studies on 
a scale-invariant extension of the SM.
There are basically two types of  scenario: one \cite{Fatelo:1994qf}-\cite{Salvio:2014soa} relies on the Coleman--Weinberg (CW) potential
\cite{Coleman:1973jx}, while the other \cite{Hur:2007uz}- \cite{Kubo:2015cna} is based on non-perturbative effects in non-abelian
gauge theory such as dynamical chiral symmetry breaking
\cite{Nambu:1960xd,Nambu:1961tp} or condensation of the gauge-invariant scalar bilinear 
\cite{Fradkin:1978dv,Abbott:1981re,Chetyrkin:1982au}.
The common thinking is that  a classically scale-invariant  physics
around TeV is  responsible for the origin of the SM scale.

Along this line of thought we have suggested 
a new model \cite{Kubo:2015cna}, in which SM singlet scalar fields $S$ are 
coupled with non-abelian gauge fields in a hidden sector.
Below a certain energy scale   the scalar fields condensate 
in the form of the bilinear, i.e. $\langle S^\dagger S \rangle$,
by a non-perturbative effect  of the hidden sector.
Because of the condensate  the Higgs portal term turns to
a Higgs mass term with a squared  mass proportional to
 $\langle S^\dagger S\rangle$.
However, this is too naive, because it is a non-perturbative
effect, and  there is a back reaction on the condensate
from the Higgs through the portal.
In \cite{Kubo:2015cna} we have proposed an effective theory 
for the condensation of scalar bilinear and investigated the vacuum structure in the self-consistent mean field approximation (SCMFA) \cite{Kunihiro:1983ej,Hatsuda:1994pi}.
Furthermore, we have introduced flavors to the scalar fields and shown that realistic DM candidates, which are the excited states above the vacuum,
 exist in the model. 
Thus, the DM  and EW  scales have the same origin.

In this paper we will study phase transitions 
at finite temperature in our model.
There will be EW and scale phase transitions.
As is well known a strong first-order  EW phase transition is important  for
baryon asymmetry in the Universe \cite{Kuzmin:1985mm}
-\cite{Dine:1992wr}.
By the scale phase transition we mean a transition between phases
with a zero and nonzero condensates of the scalar bilinear.
Note that (to the best of our knowledge)
the scale phase transition 
in a non-abelian gauge theory has not been studied
and therefore the nature of the phase transition is not known.
Since we have an effective theory for the condensation of the scalar bilinear
at hand, we will 
address this problem by means of the  effective theory.
The first sections  will be used to explain  the model as well as 
the effective theory.
We expect that 
there exists a nontrivial interplay
between the EW and scale phase transitions,
because the EW scale is created by the condensate in the hidden sector.
We will be able to  confirm this expectation in Sect.~\ref{sec:phase transition}.
Moreover, it will turn out that 
the EW and scale phase transitions can be a strong first-order
phase transition
in a certain parameter space of the model.
Section.~\ref{sum} will be devoted to a summary.

\section{The model and its effective Lagrangian}\label{models}
\noindent Our hidden sector \cite{Kubo:2015cna} consists of 
 strongly interacting $SU(N_c)$ gauge fields coupled with
 the scalar fields $S_i^{a}~
(a=1,\dots,N_c,i=1,\dots,N_f)$ in
the fundamental representation of $SU(N_c)$.
The hidden sector Lagrangian is given by
\be
{\cal L}_{\rm H} &=&-\frac{1}{2}~\mbox{tr} F^2+
([D^\mu S_i]^\dag D_\mu S_i)-
\hat{\lambda}_{S}(S_i^\dag S_i) (S_j^\dag S_j)\nn\\
& &-\hat{\lambda'}_{S}
(S_i^\dag S_j)(S_j^\dag S_i)
+\hat{\lambda}_{HS}(S_i^\dag S_i)H^\dag H,
\label{LH}
\ee
where $D_\mu S_i = \partial_\mu S_i -ig_H G_\mu S_i$, $G_\mu$ is the matrix-valued gauge field, 
the trace is taken over  the color indices, and
the SM Higgs doublet field is denoted by $H$.
The total Lagrangian is the sum of 
${\cal L}_{\rm H}$ and ${\cal L}_{\mathrm{SM}}$, where
 the scalar potential of the SM part, 
 ${\cal L}_{\mathrm{SM}}$,
 is
\be
V_{\mathrm{SM}}
&=&\lambda_H ( H^\dag H)^2.
\label{VSM}
\ee
Note that the Higgs mass term is absent.
Below a certain energy scale
the gauge coupling $g_H$ becomes so large that
the $SU(N_c)$ invariant scalar bilinear dynamically
forms a $U(N_f)$ invariant
condensate \cite{Abbott:1981re,Chetyrkin:1982au},
\be
\langle (S^\dag_i S_j)\rangle &=&
\langle ~\sum_{a=1}^{N_c} S^{a\dag}_i S^a_j~\rangle\propto \delta_{ij},
\label{condensate}
\ee 
which  breaks  classical scale invariance.  But the condensate 
(\ref{condensate})  is not 
an order parameter, because  scale invariance is broken by
scale anomaly, too \cite{Callan:1970yg}.
This hard breaking by anomaly  is only logarithmic, and it implies that
that the coupling constants depend on the energy scale
\cite{Callan:1970yg}.
Therefore,  we have assumed in \cite{Kubo:2015cna} that 
the non-perturbative  breaking
is dominant,
 so that we can ignore the scale anomaly 
 in writing down  an effective Lagrangian
to the condensation of the scalar bilinear  at the tree level.
The effective Lagrangian 
does not contain the $SU(N_c)$ gauge fields, because 
they are integrated out, while it contains
the ``constituent'' scalar fields $S_i^{a}$.
Since the effective theory
should dynamically describe  
the condensation of the scalar bilinear, which should be the origin of
the breaking of scale invariance,
the effective Lagrangian has to be invariant under
scale transformation:
 \be
 {\cal L}_{\rm eff} &=& 
 ([\partial^\mu S_i]^\dag \partial_\mu S_i)-
\lambda_{S}(S_i^\dag S_i) (S_j^\dag S_j)
-\lambda'_{S}
(S_i^\dag S_j)(S_j^\dag S_i)\nn\\
& &+\lambda_{HS}(S_i^\dag S_i)H^\dag H
-\lambda_H ( H^\dag H)^2,
\label{Leff}
\ee
where we assume that  all $\lambda$'s are positive. 
This is the most general form which is
consistent with 
the $SU(N_c)\times U(N_f)$ symmetry and the classical scale invariance,
where  the kinetic term for $H$ 
is included in ${\cal L}_{\mathrm{SM}}$.\footnote{Quantum field theory defined by (\ref{Leff})
with the kinetic term for $H$ is renormalizable 
in perturbation theory \cite{Lowenstein:1975rf}.}
That is, ${\cal L}_{\rm H}-V_{\mathrm{SM}}$ 
has the same global symmetry as  ${\cal L}_{\rm eff}$ even at the quantum level, where ${\cal L}_{\rm H}$ and $V_{\mathrm{SM}}$
are given in (\ref{LH}) and (\ref{VSM}), respectively.
Note that the couplings
$\hat{\lambda}_{S}, \hat{\lambda'}_{S}$, and $\hat{\lambda}_{HS}$ in ${\cal L}_{\rm H}$ are not the same as
$\lambda_{S}, \lambda'_{S}$, and $\lambda_{HS}$ in $ {\cal L}_{\rm eff}$,
because the latter are effective couplings which are dressed by
 hidden gluon contributions.

\section{Self-consistent mean field approximation}\label{mfaapp}
\noindent In the SCMF approximation \cite{Kunihiro:1983ej},
which has proved 
to be a successful approximation for  the   
Nambu--Jona-Lasinio theory   \cite{Nambu:1961tp},
the perturbative vacuum  is 
Bogoliubov--Valatin (BV) transformed to $ | 0_{\rm B} \rangle$, such that
\be
\langle 0_{\rm B}  | (S_i^\dag S_j)| 0_{\rm B}  \rangle &=&
f_{ij}=\langle f_{ij}\rangle +Z_{\sigma}^{1/2}\delta_{ij}\sigma +
Z_{\phi}^{1/2}t_{ji}^\alpha \phi^\alpha~,
\label{fij}
\ee
where the real mean fields $\sigma$ and $\phi^\alpha~
(\alpha=1,\dots,N_f^2-1)$ are introduced as
the excitations of the condensate 
$\langle f_{ij}\rangle$.
Here, $t^\alpha$ (normalized   as $
\mbox{Tr} (t^\alpha t^\beta)=\delta^{\alpha\beta}/2 $) are
the $SU(N_f)$ generators in the hermitian matrix representation, and
$Z_\sigma$ and $Z_\phi$ are the  wave function renormalization constants
of  a canonical  dimension $2$.
The unbroken $U(N_f)$ flavor symmetry  implies
\be
\langle f_{ij}\rangle & =&\delta_{ij} f~\mbox{and}~
\langle \sigma\rangle=\langle \phi^\alpha\rangle =0~,
\label{f-defined}
\ee
where a nonzero $\langle \sigma\rangle$
 can be  absorbed into $f$, so that we can always assume
 $\langle \sigma\rangle =0$.

In the SCMF approximation $f$ 
is determined in a self-consistent way as follows.
One  first splits up the effective Lagrangian (\ref{Leff}) into the sum, i.e.,
$\mathcal{L}_{\rm eff} =\mathcal{L}_{\rm MFA}+\mathcal{L}_{I}$,
where  $\mathcal{L}_{I}$ is normal ordered 
(i.e. $\langle 0_{\rm B}\vert \mathcal{L}_{I}\vert 0_{\rm B}\rangle =0$), and $\mathcal{L}_{\rm MFA}$ contains at most  bilinear terms of $S$ 
 which are not normal ordered. Using the Wick theorem
\be
(S^\dag_i S_j) &=&
:(S^\dag_i S_j): +f_{ij}~,~
(S^\dag_i S_j)(S^\dag_j S_i) =
:(S^\dag_i S_j)(S^\dag_j S_i) :+2f_{ij}(S^\dag_j S_i)
-|f_{ij}|^2
\ee
etc., we find
 \be
\mathcal{L}_{\rm MFA}
&=&(\del^\mu S^\dag_i\del_\mu S_i)
-M^2(S ^\dagger_i S_i)
+N_f(N_f\lambda_S+\lambda'_{S})Z_{\sigma}\sigma^2
+\frac{\lambda'_{S} }{2}Z_{\phi}\phi^\alpha \phi^\alpha \nn\\
& &-2(N_f \lambda_S+\lambda'_{S})
Z_{\sigma}^{1/2}\sigma(S ^\dagger_i S_i)
-2\lambda'_{S}Z_{\phi}^{1/2} (S^\dag_i t^\alpha_{ij} \phi^\alpha S_j)
\nn\\
&  &+\lambda_{HS}(S ^\dagger_i S_i)H^\dag H
 -\lambda_H (H^\dag H)^2,
 \label{LMFA}
\ee
where 
\be
M^2=2( N_f \lambda_S +\lambda'_{S})f
-\lambda_{HS} H^\dag H,
\label{M2}
\ee
and the linear term in $\sigma$  is suppressed because it will be cancelled against the corresponding
tad pole correction.
To the lowest order in the SCMF approximation,
the ``interacting '' part $\mathcal{L}_I$ does not contribute
to the amplitudes without external $S$'s (the mean field vacuum amplitudes).
We emphasize that, in applying  the Wick theorem,  
 only the $SU(N_c)$ invariant bilinear product
$(S^{\dag}_i S_j)=\sum_a^{N_c} S^{a\dag}_i S^a_j$
 has a non-zero  (BV transformed) vacuum expectation value.

 Given the effective Lagrangian $\mathcal{L}_{\rm MFA}$,
 we next compute an effective potential $V_{\rm MFA}$
 by integrating out the mean field fluctuations $S^a_i$,
 where the fluctuations of the SM fields including $H$ will
 be taken into account later on when discussing finite temperature effects.
We employ the $\overline{\mbox{MS}}$
scheme, 
because dimensional regularization does not break scale invariance.
To the lowest order the divergences can be removed by
renormalization of $\lambda_I~(I=H,S,HS)$, i.e.
$\lambda_I\to (\mu^2)^\epsilon (\lambda_{I}+\delta\lambda_{I})$
and also by  the shift
$f\to f+\delta f$,
where $\epsilon =(4-D)/2$, and $\mu$ is the  scale introduced in dimensional regularization.
The effective potential for $\mathcal{L}_{\rm MFA}$ can be straightforwardly 
computed :
\be
V_{\rm MFA}
&=&
M^2(S_i ^\dagger S_i)
+\lambda_H(H^\dagger H)^2-
N_f(N_f\lambda_S+\lambda'_S )f^2+\frac{N_c N_f}{32\pi^2}
M^4\ln\frac{M^2}{\Lambda_H^2}~,
\label{VMFA}
\ee
where $\Lambda_H=\mu \exp (3/4)$ is so chosen that the loop correction vanishes at
$M^2=\Lambda_H^2$.
$V_{\rm MFA}$ with a term linear in $f$ included 
but without the Higgs doublet $H$ has also been discussed in 
\cite{Coleman:1974jh,Kobayashi:1975ev,Abbott:1975bn,Bardeen:1983st}.
The classical scale invariance forbids the presence of this linear term.
To find the minimum of 
$V_{\rm MFA}$ we look for the solutions of
\be
0&=&\frac{\del}{\del S^a_i}V_{\rm MFA}
= \frac{\del}{\del f}V_{\rm MFA}
=\frac{\del}{\del H_l}V_{\rm MFA}~(l=1,2).
\label{station}
\ee
The first equation gives
$0=\langle S^{a}_i\rangle^{\dag}\langle M^2\rangle=
\langle S^{a}_i\rangle^{\dag}\langle  2(N_f\lambda_S+\lambda'_S )f-\lambda_{HS} H^\dag H\rangle $,
which has three solutions: (i) $ \langle S^{a}_i
 \rangle \neq 0~\mbox{and}~\langle M^2 \rangle =0$; 
(ii) $ \langle S^{a}_i \rangle = 0~\mbox{and}~\langle M^2 \rangle =0$; and (iii)
$ \langle S^{a}_i \rangle = 0~\mbox{and}~\langle M^2 \rangle\neq 0$.
One can easily convince oneself that the solution (i)
implies $ G =0 $
if the second and third equations in (\ref{station}) are used, where
\be
G &=& 4N_f \lambda_H \lambda_S-N_f \lambda_{HS}^2+
4 \lambda_H\lambda'_S.
\label{G}
\ee
Therefore, the solution (i) is inconsistent,
unless we use the fine-tuned relation among the coupling constants.
Next, we consider the solution (ii) and find that $ \langle S^{a}_i
 \rangle =\langle f \rangle=\langle H\rangle =0$
 with  $\langle V_{\rm MFA}\rangle =0$.
The third solution  (iii) can exist
if $G> 0$
 is satisfied, and 
 we find
\be
 |\langle H\rangle |^2
 &=&\frac{v_h^2}{2}=
 \frac{N_f\lambda_{HS}}{G}\Lambda_H^2\exp\left(  \frac{32\pi^2 \lambda_H}{N_c G}-\frac{1}{2}\right),~
 \langle f\rangle =\frac{2 \lambda_H}{N_f\lambda_{HS}} 
 |\langle H\rangle |^2,
 \label{vev1}\\
 \langle M^2\rangle &=& M_0^2=\frac{G}{N_f\lambda_{HS}}
  |\langle H\rangle |^2,~\langle V_{\rm MFA}\rangle <0.
  \label{M02}
\ee
Consequently, the solution  (iii)
presents  the true potential minimum  if $G> 0$ is satisfied.
 The Higgs mass at this level of approximation becomes
\be
m_{h0}^2 
&=& |\langle H\rangle |^2\left(
\frac{16\lambda_H^2
(N_f\lambda_S+\lambda'_S)}{G}
+\frac{N_c N_f
\lambda_{HS}^2}{8\pi^2}\right).
\label{vev2}
 \ee
In the  small $\lambda_{HS}$ limit we obtain
$m_{h0}^2 \simeq 4 \lambda_{H}| \langle H\rangle |^2
=2 \lambda_{HS}\langle f \rangle$, where
the first equation is the SM expression, 
and the second one is simply assumed in \cite{Kubo:2014ova}.
There will be a correction ($\sim 7\%$) to (\ref{vev2}) coming from the 
SM part, which will be calculated later on.

We would like to note that the effective potential 
$V_{\rm MFA}$  in (\ref{VEFF}) has
a flat direction, which corresponds to the end-point contribution
of \cite{Bardeen:1983st}: 
If $M^2=2(N_f \lambda_S+\lambda'_S)f-\lambda_{HS} 
H^\dag H=0$ is satisfied, $V_{\rm MFA}=0$ for any value
of $S^a_i$, so that (except for $S^a_i=0$) 
the $SU(N_c)$ symmetry is spontaneously broken
in this direction. The origin that $\langle V_{\rm MFA} \rangle  <0$
for the solution (iii) is the absence of a mass term in the effective Lagrangian
(\ref{Leff});  we have assumed classical scale invariance to begin with.
A mass term in (\ref{Leff}) 
would  effectively generate in 
$V_{\rm MFA}$ a term linear in $f$. This linear term can lift the 
$\langle V_{\rm MFA} \rangle$ into a positive direction
\cite{Kobayashi:1975ev,Abbott:1975bn}, while
$V_{\rm MFA}=0$ remains in  the flat direction  \cite{Bardeen:1983st}.

Finally, we would like to recall once again that we regard 
the Lagrangian (\ref{Leff}) together with our approximation method
as an effective theory for the condensation of scalar bilinear,
which takes place in the $SU(N_c)$ gauge theory described by
(\ref{LH}). That is, 
we discard fundamental problems such as
the intrinsic instability  inherent in  (\ref{Leff}) \cite{Bardeen:1983st},
because we assume that such problems are absent in 
the original theory described by (\ref{LH}).

 \section{Dark matter}\label{darkmattersec}
\noindent We are now in a position to use the effective Lagrangian 
 $\mathcal{L}_{\rm MFA}$ (\ref{LMFA}) to discuss DM.
 First, we replace $M^2$ and  the Higgs doublet 
 $H$ appearing in  $\mathcal{L}_{\rm MFA}$
 by  $M^2_0$ and $H^T=( \chi^+,~(v_h+h+i\chi^0)/\sqrt{2}  )$, respectively,
where $\chi^+$ and $\chi^0$ are the would-be Nambu--Goldstone fields,
and $M_0^2$ is given in (\ref{M02}).
The linear terms in $\sigma$ and $h$ in  $\mathcal{L}_{\rm MFA}$  should be  suppressed,
because they will be cancelled against the corresponding
tad pole corrections.
 We integrate out the constituent
scalars $S^a$ to  obtain effective 
 interactions among $\sigma,~\phi$, and the Higgs $h$,
 where $\sigma$ and $\phi$ are defined in (\ref{fij}).
Their inverse propagators should be computed 
to obtain 
their masses and the corresponding wave function renormalization constants.
Up to and including one-loop order we find:
\be
\Gamma_\phi^{\alpha\beta}(p^2)
&=&Z_{\phi}\delta^{\alpha\beta}\lambda'_{S}
\Gamma_\phi(p^2)=
Z_{\phi}\delta^{\alpha\beta}\lambda'_{S}
\left[1+
2\lambda'_{S} N_c\Gamma(p^2)\right],
\label{gamma}\\
\Gamma_\sigma(p^2)
&=&2Z_{\sigma}N_f(N_f\lambda_{S}+\lambda'_{S})\left[1+
2 N_c(N_f\lambda_{S}+\lambda'_{S})\Gamma(p^2)\right],\nn\\
\Gamma_{h\sigma}(p^2)
&=&-2Z_{\sigma}^{1/2}
 v_h\lambda_{HS}(N_f\lambda_S+\lambda'_{S})
N_f N_c ~\Gamma(p^2),\nn\\
\Gamma_h(p^2)
&=& p^2-m_{h1}^2+(v_h \lambda_{HS})^2
N_f N_c ~(\Gamma(p^2)-\Gamma(0)),\nn
\ee
with $m_{h1}^2=m_{h0}^2+\delta m_h^2$, where
$m_{h0}^2$ is given in (\ref{vev2}),
$\delta m_h^2$ is the SM correction given in (\ref{dmh}), and
\be
\Gamma(p^2) &=&
\frac{1}{16\pi^2}\left(
2-\ln\left[\frac{M_0^2}{\Lambda_H^2 \exp(-3/2)}\right]
-2(4/x-1)^{1/2}~\arctan (4/x-1)^{-1/2}\right)
\ee
with $x=p^2/M_0^2$. Note that we have included the canonical kinetic term for $H$,
but  the wave function renormalization constant for $h$ is ignored, 
which is approximately equal to one within the approximation
here.
The  DM mass is the zero of the inverse propagator, i.e.
\begin{eqnarray}
\Gamma_{\phi}^{\alpha\beta}(p^2 = {m_{\rm DM}}^2)=0,
\label{zero}
\end{eqnarray}
and $Z_{\phi}$ 
(which has a canonical  dimension  $2$) can be obtained from
\be
Z_{\phi}^{-1} &=& \left.
2(\lambda'_{S})^2 N_c (d \Gamma/d p^2)
\right|_{p^2=m_{\rm DM}^2}\nn\\
&= & \frac{2(\lambda'_S)^2 N_c}{m_{\rm DM}^{2}16\pi^2}\left(~
4 ~[y(4-y)]^{-1/2}\arctan (4/y-1)^{-1/2}-1~\right)
\ee
with $y=m_{\rm DM}^2/M_0^2$.
The Higgs and $\sigma$ masses can be similarly  obtained 
from the eigenvalues of the $h-\sigma$ mixing matrix
\be
\Gamma(p^2) &=&
\left(\begin{array}{cc}
\Gamma_h(p^2) & \Gamma_{h\sigma}(p^2)\\
\Gamma_{h\sigma}(p^2)  & \Gamma_\sigma(p^2) 
\end{array}\right).
\label{mixing}
\ee
The squared Higgs and $\sigma$ masses, $m_h^2$ and $m_\sigma^2$, 
 are zeros of $\det \Gamma(p^2)$.
 That is, the SM correction
 (\ref{dmh}) and the correction from the mixing (\ref{mixing}) 
 are included in $m_h$. 
This mixing has to be taken into
account in determining the renormalization constants,
which we will ignore in the the following discussions,
because the effect is very small (as mentioned above).
In contrast, the mixing can have a non-negligible effect on the masses.
If $m_{\rm DM},m_\sigma >2 M_0$, DM or $\sigma$ 
 would decay into two $S$'s within the framework
of the effective theory, because the effective theory cannot incorporate
confinement.
Therefore, we will consider only the parameter space with
$m_{\rm DM},~m_{\sigma}< 2 M_0$.
\begin{figure}
 \includegraphics[width=6cm]{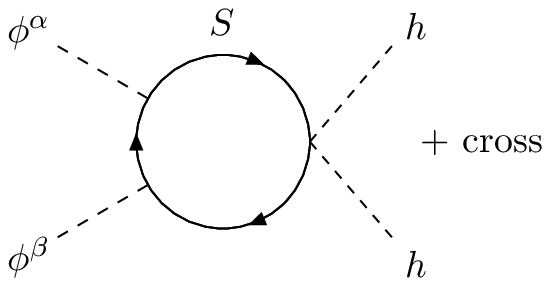}
  \includegraphics[width=12cm]{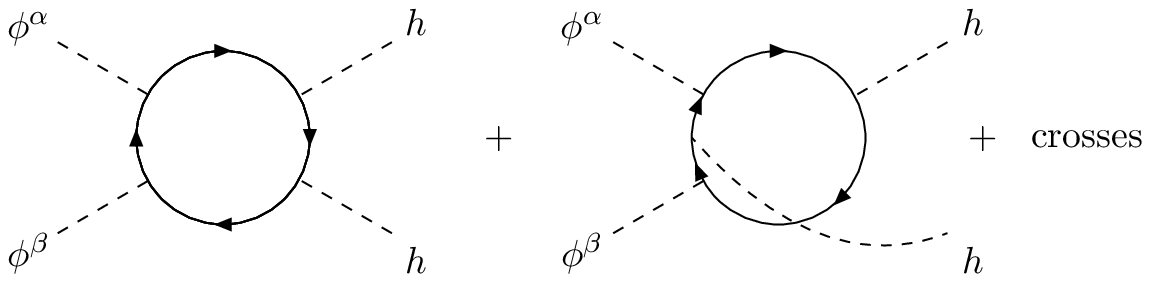}
\caption{\label{pp-hh}\footnotesize
The interaction between DM and the Higgs $h$
arises at the one-loop level.
The lower diagrams are $\sim \lambda^2_{HS} (v_h/M_0)^2$,
so that the upper diagrams are dominant,
because  $\lambda^2_{HS} (v_h/M_0)^2 <<\lambda_{HS}$
in a realistic parameter space.}
\end{figure}

  The link of  $\phi$ to the SM model
 is established through the interaction with the Higgs,
 which is generated at one-loop as shown in Fig.~\ref{pp-hh}, 
 yielding the effective couplings 
\be
\kappa_{s(t)} \delta^{\alpha\beta}
&=&  \delta^{\alpha\beta}\Gamma_{\phi^2 h^2}(M_0,m_{\rm DM},\epsilon=1(-1))~,
\label{kappa}
\ee
where
  \footnote{Since the contribution of the lower diagrams 
 in Fig.~\ref{pp-hh} is small, 
 we compute them at $p=0$, which is the
 $\epsilon$-independent term in (\ref{kappa2}).}
\be
\Gamma_{\phi^2 h^2}(M_0,m_{\rm DM}, \epsilon)
&=&
\frac{Z_\phi  N_c (\lambda'_{S})^2\lambda_{HS}}{4\pi^2 }
\left( \lambda_{HS}\frac{v_h^2}{4M_0^4}\right.\nn\\
& &\left.
-\left\{\begin{array}{c}
\frac{2}{m_{\rm DM}^2}
\left(
\frac{\arctan(4/y-1)^{-1/2}}{(4/y-1)^{-1/2} }-
\frac{\arctan(1/y-1)^{-1/2}}{(1/y-1)^{-1/2}}\right)~\mbox{for}~\epsilon=1
\\
\frac{2\arctan(4/y-1)^{-1/2}}{ M_0 m_{\rm DM}(4-y)^{1/2}}
~\mbox{for}~\epsilon=-1
\end{array}\right\}\right)~,
\label{kappa2}
\ee
$y= m^2_{\rm DM}/M_0^2$, and $v_h=246$ GeV.
We have used the s-channel $(\epsilon=1)$ momenta
$p=p'=(m_{\rm DM},{\bf 0})$ for
 DM annihilation,
because we restrict ourselves to the s-wave part of
the velocity-averaged annihilation cross section $\langle v\sigma \rangle$. 
Similarly,
we have used the t-channel $(\epsilon=-1)$ momenta
$p=-p'=(m_{\rm DM},{\bf 0})$
 for the spin-independent 
elastic cross section off the nucleon $\sigma_{SI}$. 

We obtain
\be
\langle v\sigma \rangle
&=&\frac{{1}}{32\pi m_{\rm DM}^3}~\sum_{I=W,Z,t,h}
(m_{\rm DM}^2-m_I^2)^{1/2} a_I+\mathcal{O}(v^2),\nn
\ee
where $m_W=80.4~\mbox{GeV}, m_Z=91.2~\mbox{GeV}$, and $m_t=174~\mbox{GeV}$ 
are the $W, Z$ boson and the top quark masses, respectively, and
\be
& &a_{W(Z)}= 4 (2)
[\mbox{Re}(\kappa_s)]^2\Delta_{h}^2 m_{W(Z)}^4
\left( 3+4\frac{m_{\rm DM}^4}{m_{W(Z)}^4}-4
\frac{m_{\rm DM}^2}{m_{W(Z)}^2}\right),\nn\\
& &a_t= 24 [\mbox{Re}(\kappa_s)]^2\Delta_{h}^2  
m_t^2(m_{\rm DM}^2-m_t^2),~
a_h 
=[\mbox{Re}(\kappa_s)]^2\left(
1+24 \lambda_H \Delta_{h} \frac{m_W^2}{g^2}
\right)^2.
\ee
Here, $g=0.65$ is the $SU(2)_L$ gauge coupling constant,
and $ \Delta_{h}=(4 m_{\rm DM}^2-m_h^2)^{-1}$ is 
the Higgs propagator.
The DM relic abundance \footnote{There are $(N_f^2-1)$ DM particles,
and the number of the effectively massless degrees of freedom at the freeze-out temperature is $g_*=106.75+N_f^2-1$.} is 
$\Omega \hat{h}^2 =(N_f^2-1)\times 
(Y_\infty s_0 m_{\rm DM})/(\rho_c/\hat{h}^2)$,
where $Y_\infty$ is the asymptotic value of 
the ratio $Y$ of the DM number density to entropy,
$s_0=2890\mbox{cm}^{-3}$ is the entropy density at present,
$\rho_c=
1.05 \times 10^{-5}\hat{h}^2 ~\mbox{GeV}\mbox{cm}^{-3}$ is the critical density, and
$\hat{h}$ is the dimensionless Hubble parameter.
To obtain $Y_\infty$ we solve the Boltzmann equation for $Y$.
The spin-independent 
elastic cross section off the nucleon 
$\sigma_{SI}$
is \cite{Barbieri:2006dq}
\be
\sigma_{SI}
&=&\frac{{1}}{4\pi} 
\left(\frac{\kappa_t\hat{r} m_N ^2}{m_{\rm DM}m_h^2}
\right)^2
\left(\frac{m_{\rm DM}}{m_N+m_{\rm DM}}
\right)^2,\nn
\ee
where $\kappa_t$ is given in
 (\ref{kappa}), $m_N$ is the nucleon mass, and
$\hat{r}\sim 0.3$ stems from the nucleonic matrix element 
\cite{Ellis:2000ds}.
In \cite{Kubo:2015cna} we have shown that 
there is a parameter space in the model with various
$N_f$ and $N_c$ in which 
the DM mass is of $O(1)$  TeV and 
$\sigma_{SI}$  and $\Omega \hat{h}^2$
are, respectively, consistent with the recent experimental
measurements in \cite{Akerib:2013tjd} and \cite{Planck:2015xua}.

 \section{Phase transitions at finite temperature}\label{sec:phase transition}
\noindent At a certain finite temperature the condensation of the scalar bilinear
will be dissolved, and above that temperature 
the EW symmetry will be restored.
The nature of the  EW symmetry breaking is crucial  for
baryon asymmetry in the Universe 
\cite{Kuzmin:1985mm,Klinkhamer:1984di,Arnold:1987mh,Shaposhnikov:1986jp}. 
Here we  investigate how the scale  and 
 EW symmetry breakings disappear
as temperature increases from a low temperature.
\footnote{EW baryogenesis in a scale-invariant extension
of the two-Higgs doublet model has been analyzed in 
\cite{Farzinnia:2014yqa,Fuyuto:2015jha,Sannino:2015wka}.}
To this end,
we integrate out the quantum
fluctuations at  finite temperature within the framework of the 
effective theory in the mean field approximation. As a result we obtain 
an effective potential at  finite temperature consisting of four components 
\cite{Anderson:1991zb,Carrington:1991hz,Dine:1991ck,Dine:1992wr}:
\begin{align}
V_{\rm eff}(f,h,T) &=V_{\rm MFA}(f,h)+
V_{\rm CW}(h) +V_{\rm FT}(f,h,T)+V_{\rm RING}(h,T)~, \displaybreak[0]
\label{VEFF}
\end{align}
where $V_{\rm MFA}(f,h)$ is the effective potential
given in (\ref{VMFA}) with $S_i^a=0$ and $H$ replaced
by $h/\sqrt{2}$, and $f$ (the condensate) is defined in (\ref{f-defined}).
Further, 
 $V_{\rm CW}(h)$ and  $V_{\rm FT}(f,h,T)$ are  the one-loop  contributions  at zero and finite temperature $T$, respectively, and $V_{\rm RING}$ is the ring contribution. 
The Coleman--Weinberg potential $V_{\rm CW}(h)$ is normalized such that
\be
V_{\rm CW}(h=v_h) &=&0,~
\frac{d V_{\rm CW}(h) }{d h}\left|_{h=v_h}\right.=0,
\label{normalization}
\ee
where we use $v_h=\langle h\rangle|_{T=0}=246$ GeV.
This normalization ensures that the  potential $V_{\rm CW}(h)$
does not change $v_h$ given in (\ref{vev1}) obtained from $V_{\rm MFA}(f,h)$.
It  can be explicitly written as
\be
V_{\rm CW}(h) &=&
C_0 (h^4-v_h^4)+\frac{1}{64 \pi^2}
\left[ ~6 \tilde{m}_W^4 \ln (\tilde{m}_W^2/m_W^2)+
3 \tilde{m}_Z^4 \ln (\tilde{m}_Z^2/m_Z^2)\right.\nn\\
 & &\left.+ \tilde{m}_h^4 \ln (\tilde{m}_h^2/m_h^2)
  -12\tilde{m}_t^4 \ln (\tilde{m}_t^2/m_t^2)~\right],
  \label{VCW}
\ee
where
\be
C_0 &\simeq &-\frac{1}{64 \pi^2 v_h^4}\left(3m_W^4+(3/2)m_Z^4+(3/4) m_h^4-
6 m_t^4\right)~,\label{C0}\\
\tilde{m}_W^2 &=&(m_W/v_h)^2h^2,~
\tilde{m}_Z^2=(m_Z/v_h)^2 h^2,~
\tilde{m}_t^2 =(m_t/v_h)^2 h^2,\nn\\
\tilde{m}_h^2 &=&
3\lambda_H h^2
+\frac{\lambda_{HS}}{64\pi^2}\left\{7 N_c N_f \lambda_{HS}h^2
-4 f N_c N_f (N_f\lambda_{S}+\lambda'_{S})\right.\nn\\
& &\left. -2 N_c N_f \left[-3 \lambda_{HS} h^2
+4f (N_f \lambda_S+\lambda'_S)\right]
  \ln \frac{4 f (N_f\lambda_S+\lambda'_S)
-\lambda_{HS}h^2}{2 \Lambda_H^2}\right\}.
\label{mh2}
\ee
We work in the Landau gauge, in which the Faddeev--Popov ghost fields are massless
even at finite temperature, so that they  do not contribute to 
$V_{\rm eff}$. The would-be NG bosons are massless
only at the potential minimum. But we have neglected their contributions
in (\ref{VCW}), because they are negligibly small.
The tedious expression for $\tilde{m}^2_h$ comes from the fact that
the Higgs mass is generated from the condensation
of the scalar bilinear: 
it is the second derivative of $V_{\rm MFA}$ in (\ref{VMFA})  with respect to $h$.
Note that $V_{\rm CW}(h) $  
contributes to the Higgs mass  correction
\footnote{The Higgs mass correction and also $C_0$ in (\ref{C0})
look more complicated if we use the Higgs mass (\ref{mh2}).
So, the term $\propto m_h^4$ in (\ref{C0}) and (\ref{dmh}) 
is only an approximate expression.}
\be
\delta m_h^2 &\simeq &-16 C_0v_h^2,
\label{dmh}
\ee
which is about $7$\% in $m_h$.
We follow \cite{Carrington:1991hz} and find
\be
V_{\rm FT}(f,h,T)
&=&\frac{T^4}{2\pi^2}\left(
2N_c N_f J_B(\tilde{M}^2(T) /T^2)+
J_B(\tilde{m}_h^2(T) /T^2)\right.\nn\\
& &\left. +6J_B(\tilde{m}_W^2 /T^2)+
3J_B(\tilde{m}_Z^2 /T^2)-12J_F(\tilde{m}_t^2/T^2)\right)
\label{VFT},
\label{VT}
\ee
where the thermal masses are 
\be
\tilde{M}^2(T) 
&=& M^2+\frac{T^2}{6}\left(~ (N_c N_f +1)\lambda_S +(N_f+N_c)\lambda'_S
-\lambda_{HS}~\right),\\
\tilde{m}_h^2(T) 
&=& \tilde{m}_h^2+
\frac{T^2}{12}\left(\frac{9}{4}g^2+
\frac{3}{4}g'^2+3 y_t^2+6\lambda_H -N_c N_f \lambda_{HS}\right),
\ee
the  coupling constants
$g=0.65$, $g'=0.36$, and $y_t=1.0$ stand for 
the $SU(2)_L$, $U(1)_Y$ gauge coupling constants and
the top Yukawa coupling constant, respectively, and $M$ is defined in (\ref{M2}) with $H^\dag H=h^2/2$.
The thermal functions $J_B$ and $J_F$ are defined as
\be
J_B(r^2) &=&\int_0^\infty dx x^2 \ln
\left(1-e^{-\sqrt{x^2+r^2} } \right)\nn\\
&\simeq &
-\frac{\pi^4}{45}+\frac{\pi^2}{12}r^2
-\frac{\pi}{6}r^{3}-\frac{r^4}{32}
\left[\ln (r^2 /16\pi^2)+2\gamma_E-\frac{3}{2}   \right]
\mbox{for}~r^2\lsim 2,
\label{JB}\\
J_F(r^2) &=&\int_0^\infty dx x^2 
\ln\left(1+e^{-\sqrt{x^2+r^2} } \right)\nn\\
&\simeq &
\frac{7\pi^4}{360}-\frac{\pi^2}{24}r^2
-\frac{r^4}{32}\left[\ln (r^2 /\pi^2)+2\gamma_E-\frac{3}{2}
   \right] \mbox{for}~r^2\lsim 2.
   \label{JF}
\ee
In the actual calculations we employ the idea 
\cite{Funakubo:2009eg}
for  approximating  the thermal functions
as
\be
J_{B(F)}(r^2)
&\simeq &\
\exp (-r)\sum_{n=0}^{40} c_n^{B(F)} r^n.
\ee
Finally, the ring contribution from the gauge bosons is \cite{Carrington:1991hz}
\be
V_{\rm RING}
&&=
-\frac{T}{12\pi}\left(
2 a_g^{3/2}+\frac{1}{2\sqrt{2}}\left(a_g+c_g-[(a_g-c_g)^2+4 b_g^2]^{1/2}\right)^{3/2}\right.\nn\\
&&\left.
+\frac{1}{2\sqrt{2}}\left(a_g+c_g+[(a_g-c_g)^2+4 b_g^2]^{1/2}\right)^{3/2}-\frac{1}{4}[g^2 h^2]^{3/2}
-\frac{1}{8}[(g^2+g'^2) h^2]^{3/2}\right),
\ee
where
\be
a_g &=&\frac{1}{4}g^2 h^2+\frac{11}{6}g^2 T^2,~
b_g = -\frac{1}{4}g g' h^2~,~
c_g= \frac{1}{4}g'^2 h^2+\frac{11}{6}g'^2 T^2.
\ee
The critical temperatures of the scale phase and EW phase transitions
(which we denote by $T_{\rm S}$ and $T_{\rm EW}$, respectively) can
be different. If $T_{\rm S}$ and $T_{\rm EW}$ are distant from each other,
two phase transitions cannot influence each other much.
In the case that they are close or equal, i.e.
$T_{\rm C}\equiv T_{\rm S}=T_{\rm EW}$,
two phase transitions  can influence each other.
In fact, depending on the choice of the parameter values,
these different cases can be realized in our model.
Below we consider some  representative examples.

\begin{figure}
  \includegraphics[width=8cm]{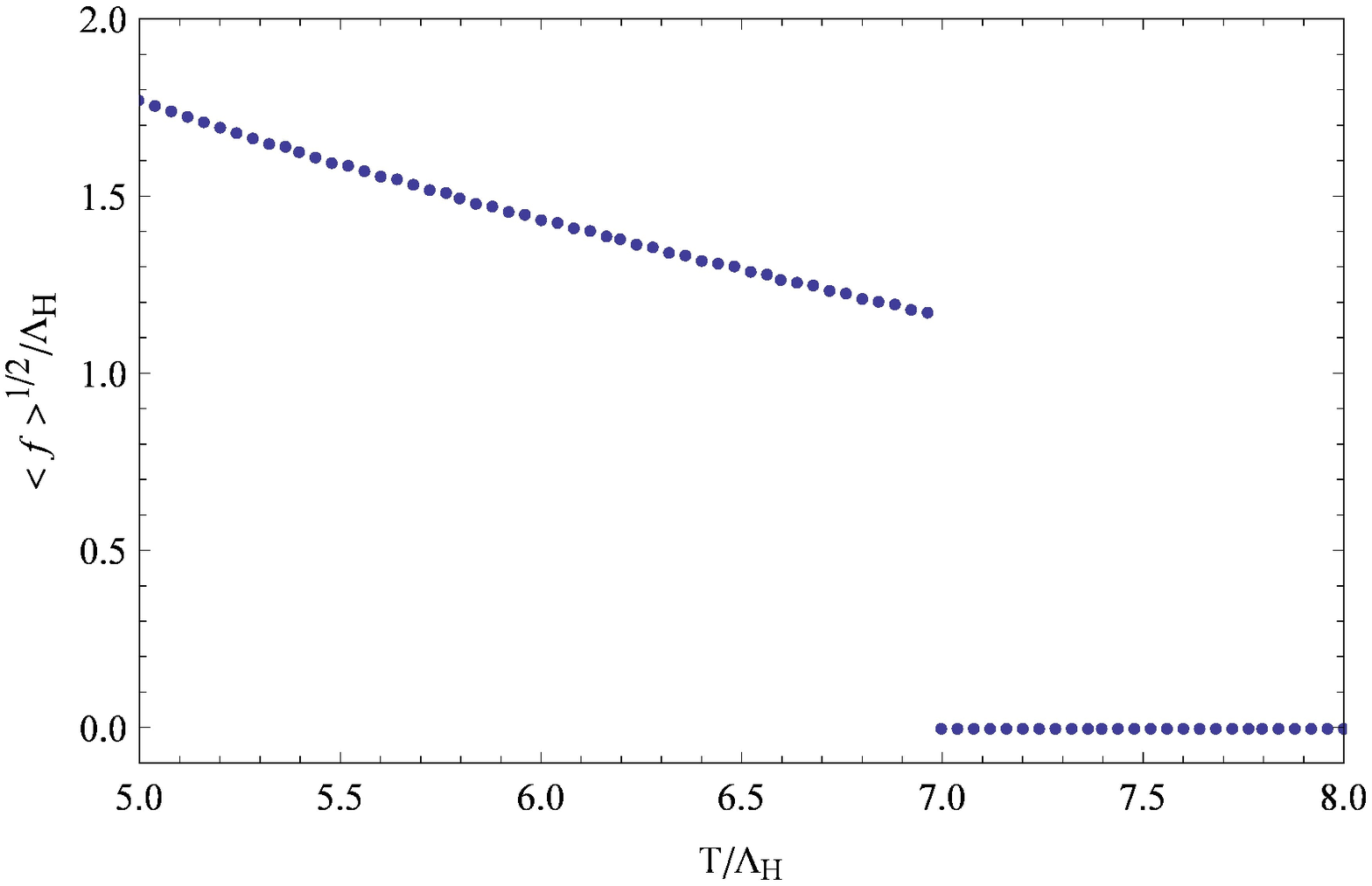} 
  \hspace{0.2cm}
  \includegraphics[width=8cm]{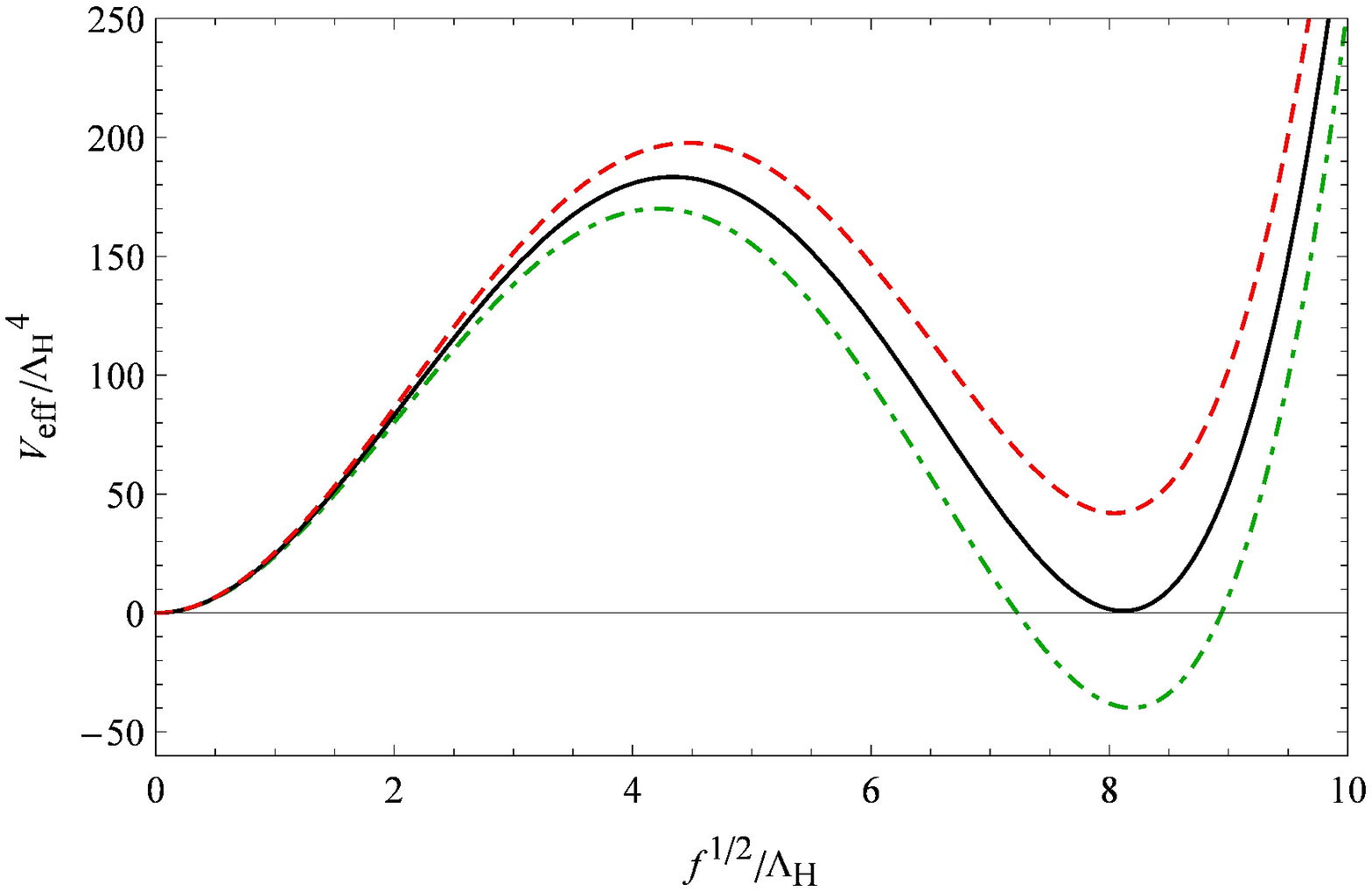}
\caption{\label{f-V}\footnotesize
 Left: The scale  phase transition for case (i), in which 
the hidden sector is disconnected from the SM. 
The (dimensionless) critical temperature is $T_{\rm S}/\Lambda_H
\simeq 7.0$. Right: The (dimensionless) potential 
$V_{\rm eff}/\Lambda_H^4 $ against 
$f^{1/2}/\Lambda_H$ for 
$T/\Lambda_H=7.1~(\mbox{red dashed}),~T_{\rm S}/\Lambda_H~(\mbox{black}), ~
6.9~(\mbox{green dash-dotted})$. The potential energy density at the origin is subtracted 
from $V_{\rm eff}$ so that the form
of the potential for different temperatures can be compared.
}
\end{figure}

\noindent
{\bf (i) Scale phase transition  with $N_f=1,~N_c=6$}\\
 First we consider the case with $\lambda_{HS}=0$,
 i.e., no connection between the hidden sector and the SM sector.
We choose:
\be
N_f &=& 1,~N_c=6,~\lambda_{S}+ \lambda'_{S}=2.083,
 \label{input1}
 \ee
 where we will use the same $N_f$ and $N_c$ as well as
 the same parameter values for $ \lambda_{S}$ and $ \lambda'_{S}$ when discussing case (ii)
 with the SM connected.
 (If $N_f=1$,
only the linear combination $ \lambda_{S}+ \lambda'_{S}$
is an independent coupling.)
In Fig.~\ref{f-V} (left) we show $\langle  f \rangle^{1/2}/T$
against $T/\Lambda_H$. 
We see from the figure that 
the scale phase transition 
is  first order with $T_{\rm S}/\Lambda_H \simeq 7.0$.
The right panel shows the form of the potential for
$T/\Lambda_H=7.1~(\mbox{red dashed})$, $T_{\rm S}/\Lambda_H~
(\mbox{black})$, $6.9~(\mbox {green dash-dotted})$.
As we will see below, the strong first-order  scale  phase transition
in the hidden sector can infect the EW phase transition.

The existence of the first-order phase transition observed here,
was predicted in \cite{Bardeen:1983st}. In our analysis we have 
assumed (and will throughout  assume) that $\langle S_i^a\rangle=0$.
However, within the framework of the effective theory
(even if we assume classical scale invariance),
there is no reason to prefer $\langle f\rangle=\langle S_i^a\rangle=0$
to the flat direction with $\langle S_i^a\rangle \neq0$ \cite{Bardeen:1983st} 
(mentioned at the end of Sect. III) at $T > T_{\rm S}$.
We  discard this problem here, because we assume that
the local $SU(N_c)$ gauge symmetry of (\ref{LH}) remains unbroken
even at $T > T_{\rm S}$.

\begin{figure}
 \includegraphics[width=8cm]{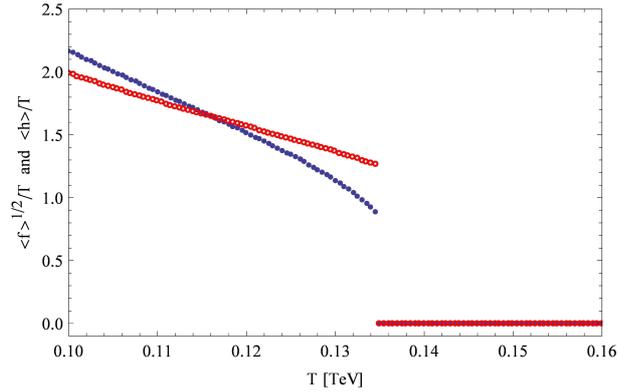}
\caption{\label{T-f-h-noDM}\footnotesize
The scale and EW phase transitions  for case (ii)
with the critical temperature
$T_{C\rm }\equiv T_{\rm S}=T_{\rm EW}\simeq 0.135$ TeV. 
The  phase transitions are both  of a strong first order.
The red circles stand for  $\langle  f \rangle^{1/2}/T$ and
the blue points are for $\langle  h \rangle/T$.
}
\end{figure}

\noindent
{\bf(ii) Scale and EW phase transitions at $T_{\rm C}\equiv T_{\rm S} = T_{\rm EW}$ }\\
 Now we couple the hidden sector with the SM sector.
 We use the same parameter values as those  given in (\ref{input1})
 along with
 \be
 \lambda_{HS} &=& 0.296, ~ \lambda_H = 0.208.
 \label{LHS}
 \ee
 The input parameters (\ref{input1}) with (\ref{LHS}) yield
$M= 0.410~\mbox{TeV},
~ m_\sigma =0.796~\mbox{TeV},~
\Lambda_H=0.019~\mbox{TeV}$, and $m_h=0.125$ TeV.
\footnote{Due to a relatively large $\lambda_{HS}$ 
there is a relatively large
mixing between $\sigma$ and the Higgs $h$
with a mixing angel of $\sim 0.2$, which is still consistent with 
the LHC constraint 
at $95\%$ CL \cite{Aad:2015zhl}. This mixing has a negative effect
on $m_h$, leading  to a large $\lambda_H$. }
In Fig.~\ref{T-f-h-noDM} we show $\langle  f \rangle^{1/2}/T$ (red)
and  $\langle  h \rangle/T$ (blue) against 
$T$, and we can see that the scale and EW phase transitions 
occur at the same critical temperature
$T_{\rm C}\equiv T_{\rm S} =T_{\rm EW}\simeq 
0.135$ TeV, where the dimensionless critical temperature
$T_{\rm C}/\Lambda_H \simeq 7.0$ is basically 
the same as that of case (i) with the SM decoupled.
This shows that the strong first-order scale phase transition 
in the hidden sector can indeed  infect the EW phase transition.

\begin{figure}
 \includegraphics[width=8cm]{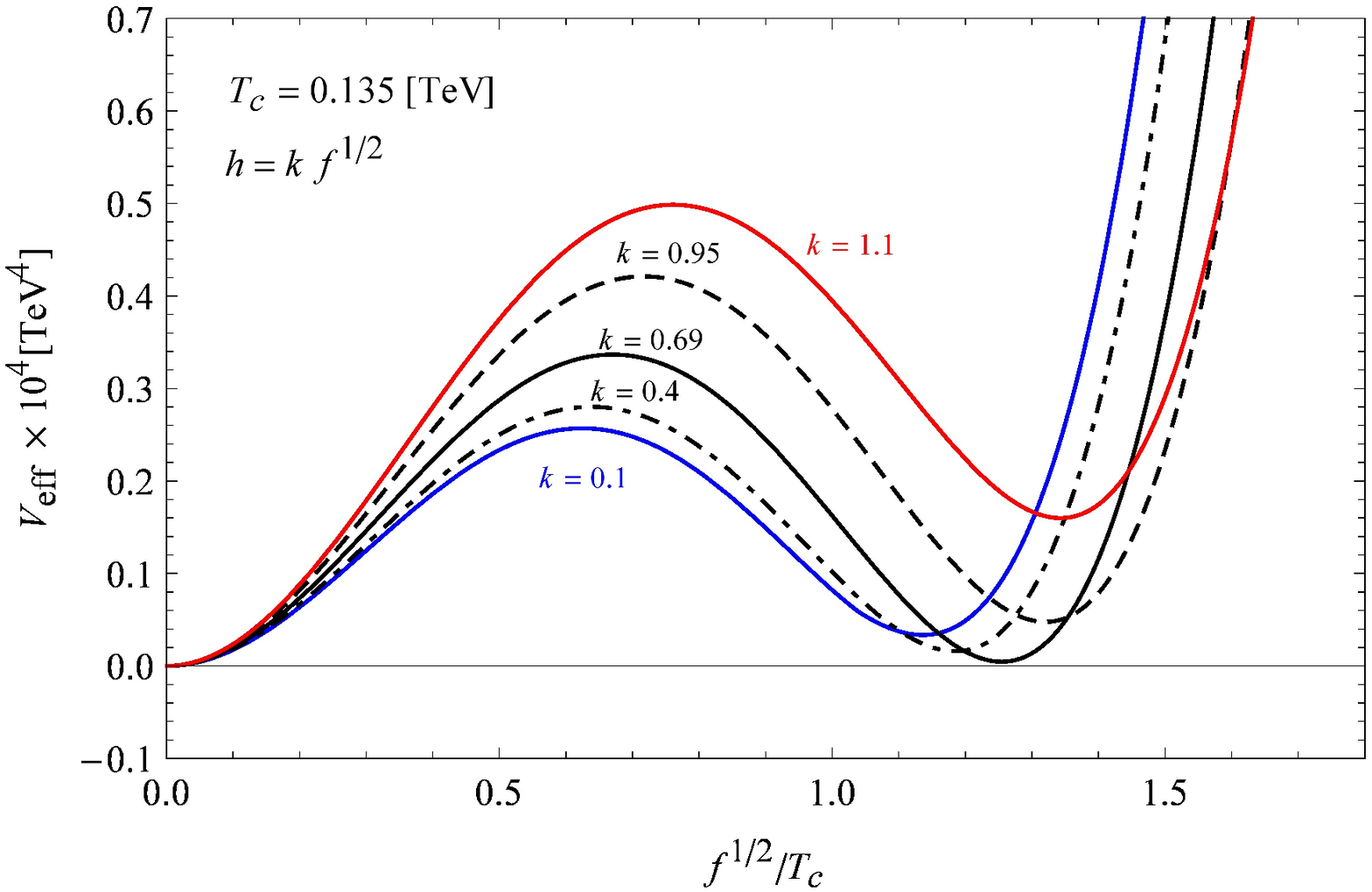}
  \includegraphics[width=8cm]{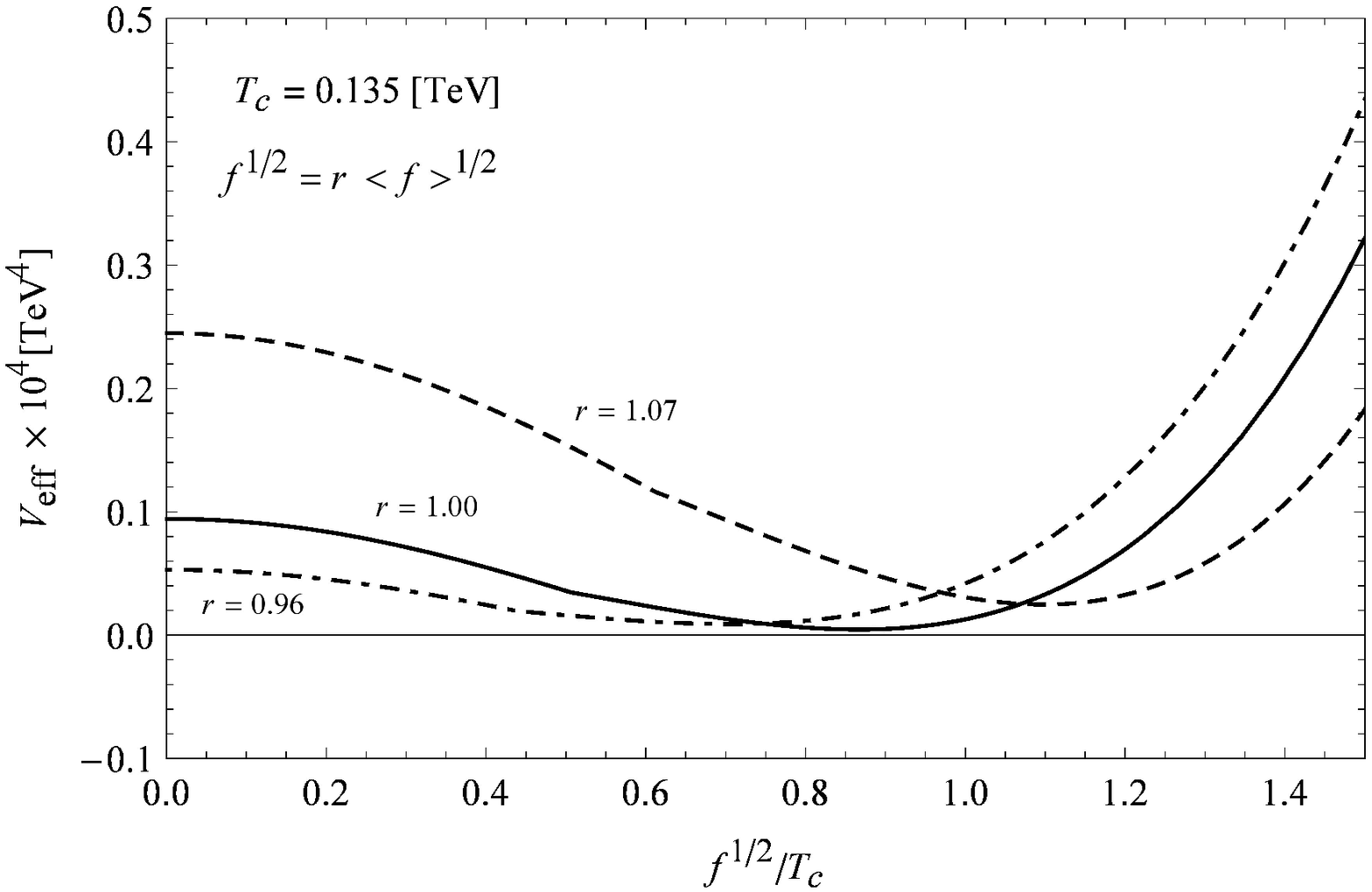}
\caption{\label{t-V-noDM}\footnotesize
The form of the potential at $T=T_{\rm C}$ for the case (ii), where
the potential energy density at the origin is subtracted 
from $V_{\rm eff}$.
Left:  
The potential as a function of $f^{1/2}/T_{\rm C}$  
on the line $h= k f^{1/2}$ in  the $f^{1/2}$--$h$ plane
with $k=1.1~(\mbox{red}),~
k=0.95~(\mbox{black dashed}),~
k=0.69~(\mbox{black}),~
k=0.4~(\mbox{black dash-dotted})$, and 
$k=0.1~(\mbox{blue})$.
Right: The potential 
as a function of $h/T_{\rm C}$ for  
$f^{1/2}$ fixed at 
$1.07 \langle f\rangle ^{1/2}\simeq 1.34 T_{\rm C}~
(\mbox{dashed}),~1.00 \langle f\rangle ^{1/2}\simeq 1.25 T_{\rm C} ~(\mbox{black})$, and $0.96 \langle f\rangle ^{1/2}\simeq 1.20 T_{\rm C}~(\mbox{dash-dotted})$.
The curve with $k= 0.69$ (left) and that with $r=1.0$ (right) 
on the potential surface
go through  the nontrivial  potential minimum.
}
\end{figure}

\begin{figure}
 \includegraphics[width=7cm]{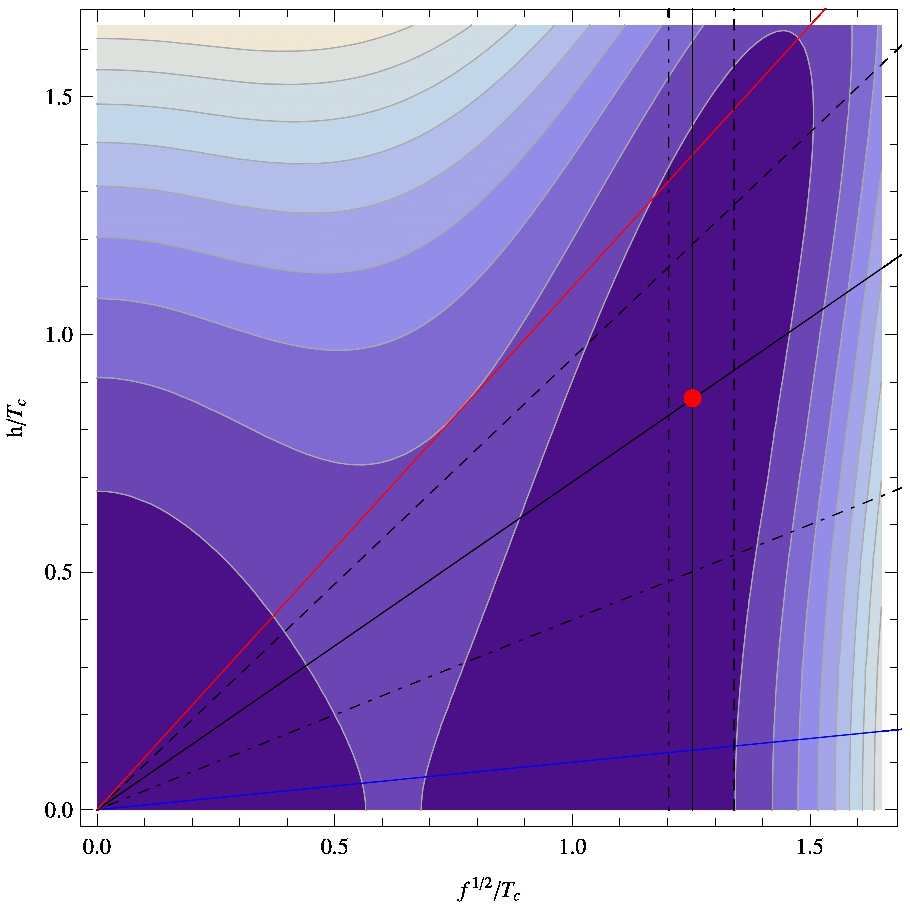}
 \includegraphics[width=1cm]{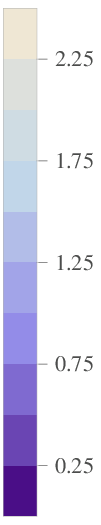}
   \includegraphics[width=7cm]{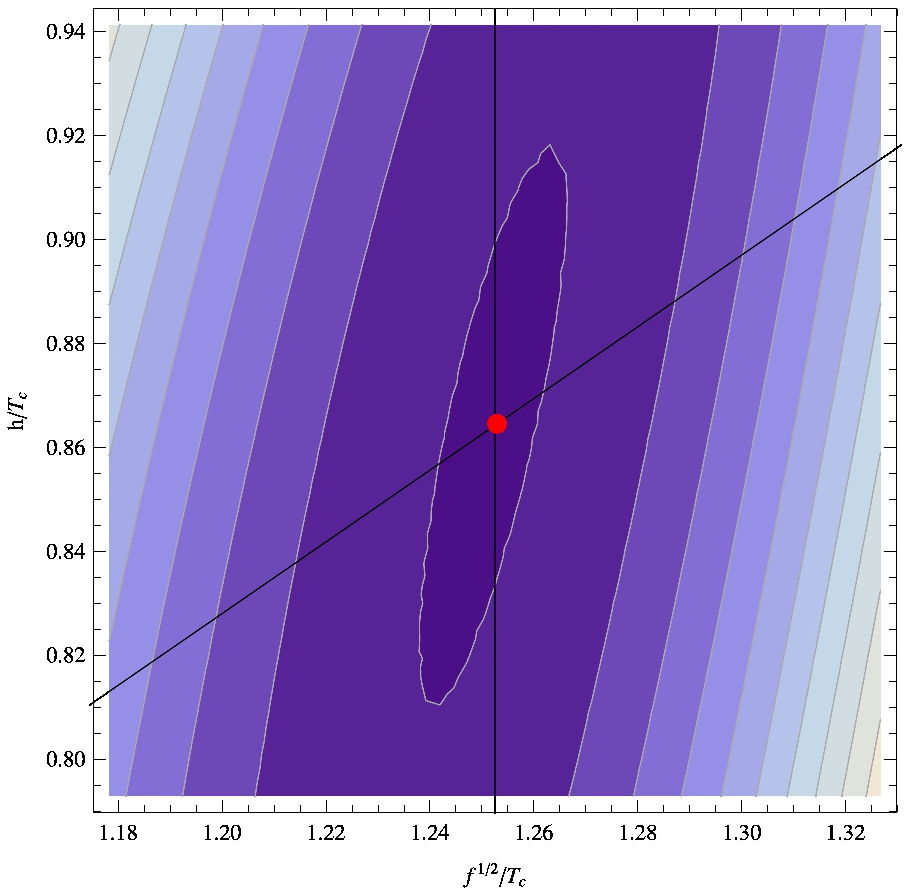}
\includegraphics[width=1cm]{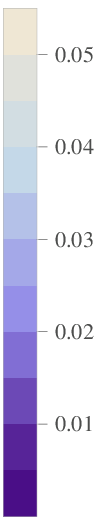}
\caption{\label{lines}\footnotesize
The lines in the $f^{1/2}/T_{\rm C}$--$h/T_{\rm C}$ plane
on which the potential values are computed and plotted in
Fig.~\ref{t-V-noDM}.
Two black lines go through  the nontrivial  potential minimum as
one can see from Fig.~\ref{t-V-noDM}.
The intersection of these two black solid lines in Fig.~\ref{lines}
 is the location of the nontrivial  potential minimum 
at $T=T_{\rm C}$, which is marked with a red point.
The darker  the color, the deeper the depth of the potential.
}
\end{figure}

We next show the form of the potential at
$T=T_{\rm C}$. 
The curves  in Fig.~\ref{t-V-noDM} (left)  are the intersections of the potential
$V_{\rm eff}$ with the surfaces defined by
\be
0 &=& h- k f^{1/2}
\label{surface}
\ee
for $k=1.1~(\mbox{red}),~
k=0.95~(\mbox{black dashed}),~
k=0.69~(\mbox{black}),~
k=0.4~(\mbox{black dash-dotted})$ and 
$k=0.1~(\mbox{blue})$,
where their intersections with 
the $f^{1/2}/T_{\rm C}-h/T_{\rm C}$ plane
are shown in Fig.~\ref{lines}.
That is, Fig.~\ref{t-V-noDM} (left) shows  the potential values on the
inclined lines in Fig.~\ref{lines} as a function of $f^{1/2}/T_{\rm C}$.
The potential minimum for $T=T_{\rm C}$ is located at the origin and
at $f^{1/2}/T_{\rm C}\simeq 1.25$ with $k\simeq 0.69$.
Since $T_{\rm C}\simeq 0.135$ TeV we obtain
$\langle f\rangle ^{1/2}\simeq 0.169$ TeV 
and $\langle h \rangle \simeq 0.117$ TeV.
Figure~\ref{t-V-noDM} (right) shows the potential 
as a function of $h/T_{\rm C}$ for  
$f^{1/2}$ fixed at 
$1.07 \langle f\rangle ^{1/2}\simeq 1.34 T_{\rm C}~
(\mbox{dashed}),~1.00 \langle f\rangle ^{1/2}\simeq 1.25 T_{\rm C} ~(\mbox{black})$, and $0.96 \langle f\rangle ^{1/2}\simeq 1.20 T_{\rm C}~(\mbox{dash-dotted})$, where these fixed values define
 the vertical lines shown in Fig.~\ref{lines}.
 The intersection of the two black solid lines in Fig.~\ref{lines} is the location
 of the potential minimum (other than the origin) at $T=T_{\rm C}$,
 which is marked with a red point.
 We have computed the potential  not only on the lines
 shown in Fig.~\ref{lines}, but also for the range 
 $0< f^{1/2}/T_{\rm C} <15,~0<h /T_{\rm C} <15$ and 
 found that there is no  other point for a minimum in this range.


\begin{figure}
  \includegraphics[width=8cm]{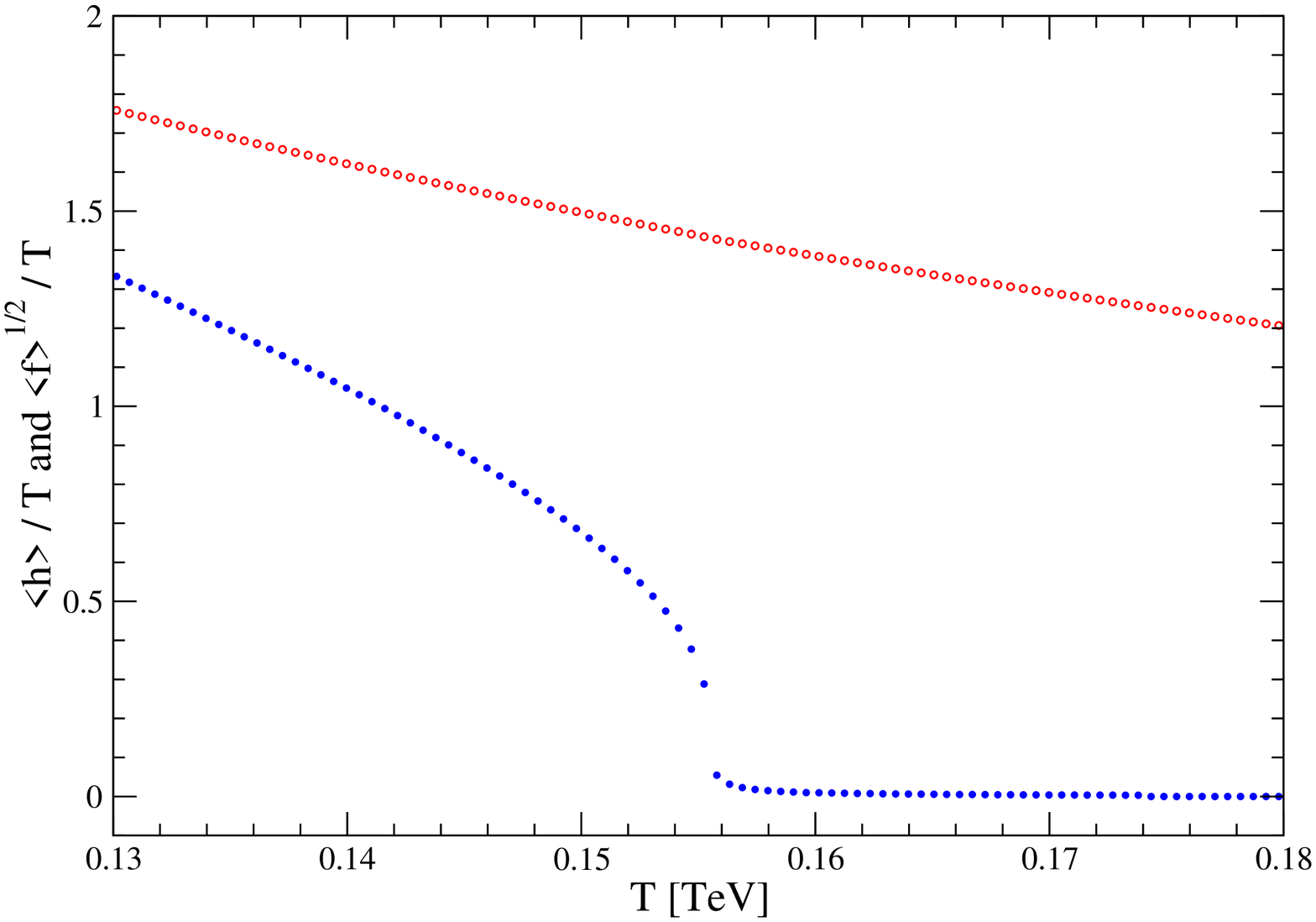} \hspace{0.2cm}\includegraphics[width=8cm]{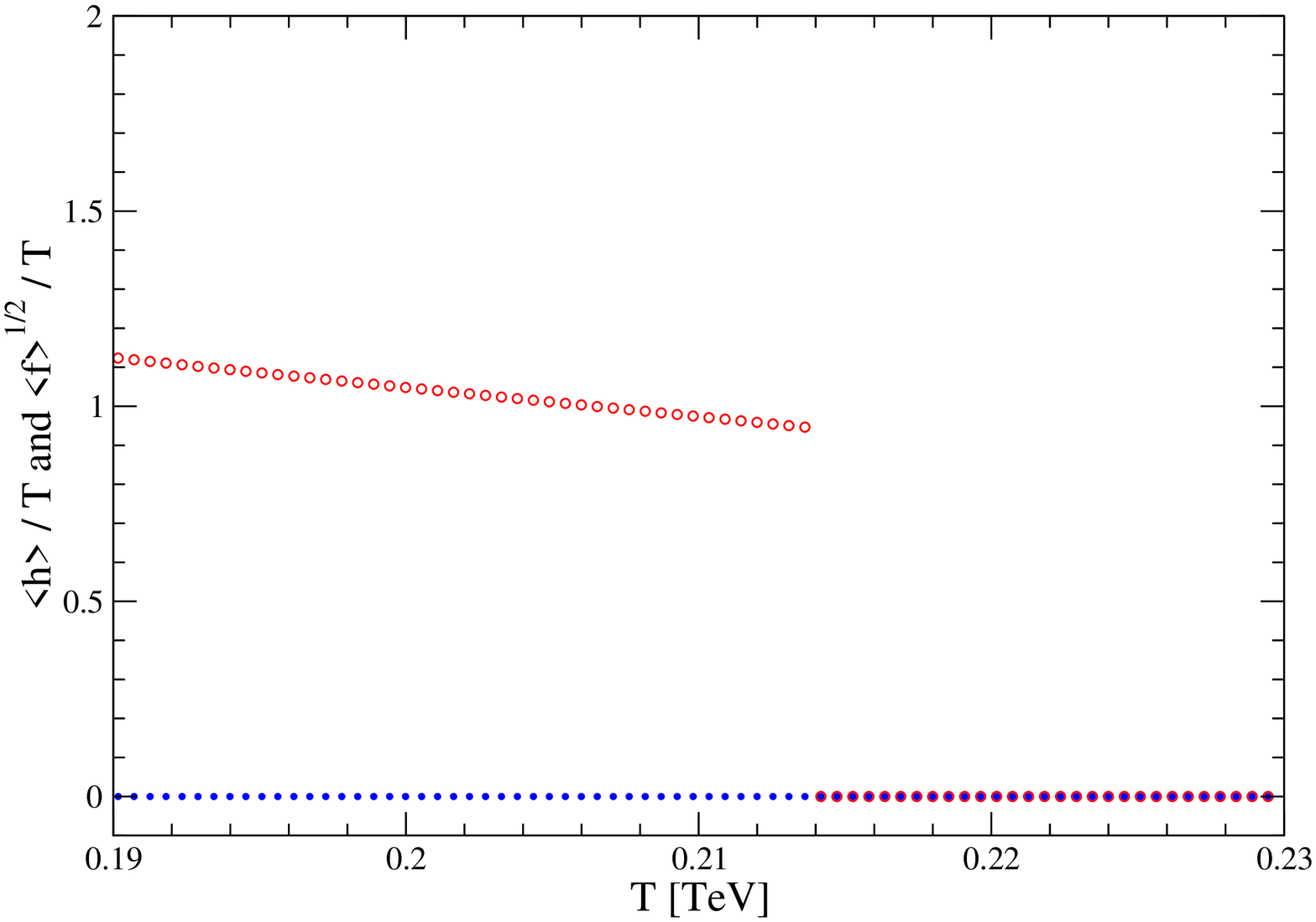}
\caption{\label{T-f-h-DM}\footnotesize
The scale  and EW phase transitions for case (iii), in which 
$T_{\rm S} > T_{\rm EW}$ is realized. 
The red circles stand for $\langle f \rangle^{1/2}/T$, while the blue
points are for  $\langle h\rangle/T$. 
The difference in the two figures is the temperature interval.
The critical temperatures are, respectively,  $T_{\rm EW}\simeq 0.155$ TeV and $T_{\rm S}
\simeq 0.214$ TeV. }
\end{figure}

\noindent
{\bf(iii) Scale and EW phase transitions with $T_{\rm S} \gsim T_{\rm EW}$ }\\
 The third example is $N_f=2$ and $N_c=6$ along with
\be
\lambda_{S}&=& 0.165, \lambda'_{S}=2.295,~
 \lambda_{HS}=0.086, ~\lambda_H=0.155.
 \ee
These input parameters yield
$M= 0.533~\mbox{TeV},
~m_{\rm DM}= 0.676~\mbox{TeV},
~ m_\sigma =0.989~\mbox{TeV},~
\Lambda_H=0.055~\mbox{TeV},~
\Omega \hat{h}^2 = 0.119$, and  $\sigma_{SI}=5.76
\times 10^{-45}~\mbox{cm}^{2}$.
In Fig.~\ref{T-f-h-DM} we show $\langle  f \rangle^{1/2}/T$ (red circles)
and  $\langle  h \rangle/T$ (blue) against 
$T$.
For the left figure the temperature $T$ varies between 
$0.13$ TeV and $0.18$ TeV, while 
$ 0.19~\mbox{TeV}~\lsim T\lsim 0.23 ~\mbox{TeV}~$ for the right figure.
We see from these figures that the critical temperatures are, respectively,
$T_{\rm EW}\simeq 0.155$ TeV and $T_{\rm S}\simeq 0.214$ TeV, and that the 
nature of the two phase transitions are
different: The scale phase transition is clearly first order,
while the nature of the  EW phase transition is indefinite.


We would like to emphasize that our results are based on the 
effective theory approach.
A more accurate calculation based on lattice simulation could alter the result.
If our observation here turns out to be correct,
the EW scalegenesis from the  
condensation of the scalar bilinear in a hidden sector
may be an alternative way to 
realize a strong first order EW phase transition.

\section{Summary}\label{sum}
\noindent We have considered the SM without the Higgs mass term, which is
coupled through a Higgs portal term, the last term of (\ref{LH}), with a  
classically scale invariant  hidden sector.
The hidden sector is an SM-singlet and described by 
an $SU(N_c)$ gauge theory
with $N_f$ scalar fields.
At lower energies
the hidden sector becomes  strongly interacting, 
and consequently the gauge-invariant scalar bilinear forms 
a  condensate (\ref{condensate}),
thereby violating scale invariance and dynamically creating a robust energy scale.
This scale is transmitted through the  Higgs portal term to the SM sector, realizing  EW scalegenesis.
Moreover, the excitation of the condensate can be identified 
with the DM degrees of freedom, which are consistent with
the present experimental observations \cite{Kubo:2015cna}.

The nature of the scale phase transition 
in a non-abelian gauge theory is not yet known.
By the scale phase transition we mean a transition between phases
with a zero and nonzero condensates of the scalar bilinear.
We have addressed this problem by means of an effective theory
for the condensation of the scalar bilinear.
Since the EW scale is (indirectly) created in the hidden sector,
it is expected that 
there exists a nontrivial interplay
between the EW and scale phase transitions. 
We have indeed confirmed this expectation
and found that there exists a parameter space 
in our model in which both 
the  EW and scale phase transitions can be 
a strong first-order phase transition.
This is not the final conclusion, because
 our result is based on the  mean field approximation
in the  effective theory.
A more accurate calculation could change this result.
It is well known that a strong first-oder phase transition 
in the early Universe
can produce gravitational wave background 
\cite{Witten:1984rs,Hogan:1986qda}, which could be observed 
by future experiments such as the Evolved Laser Interferometer Space Antenna (eLISA)
experiment \cite{AmaroSeoane:2012km}. 
In our scenario there can exist two strong first-oder phase transitions,
whose critical temperatures lie close to each other.

The nature of the  EW symmetry breaking is crucial  for
baryon asymmetry in the Universe 
\cite{Kuzmin:1985mm,Klinkhamer:1984di,Arnold:1987mh,Shaposhnikov:1986jp}. For a successful EW baryogenesis, there have to exist CP phases other than
that of the SM. Unfortunately, there is no such phase in our model
as it stands. We will come to an extension of the model
so as to realize a successful EW baryogenesis elsewhere.

\vspace{0.3cm}
\noindent{\bf Acknowledgements:} \\
We thank K.~S.~Lim,  M.~Lindner  and S.~Takeda for useful
discussions. We also thank the theory group of the Max-Planck-Institut
f\"ur  Kernphysik for their kind hospitality.
The work of M.~Y is supported by 
a Grant-in-Aid for JSPS Fellows (No. 25-5332).

\end{document}